\documentclass{KapProc} 
\kluwerbib
\let\lcitebracket(
\let\rcitebracket)
\usepackage[dvips]{graphics,epsfig}

\newcommand{\gtsim}{\mbox{{\raisebox{-0.4ex}{$\stackrel{>}{{\scriptstyle\sim}}
$}}}}

\begin{document}


\articletitle{The radio galaxy $K-$\lowercase{$z$} relation to \lowercase{$z$}$\, \sim 4.5$ }
\author{Matt J.\,Jarvis$^{1,2}$, Steve Rawlings$^2$, Steve Eales$^3$\\
Katherine M.\,Blundell$^2$, Chris J.\,Willott$^2$} 
\affil{$^1$Sterrewacht Leiden, The Netherlands\\
$^2$University of Oxford, UK\\
$^3$University of Wales College of Cardiff, UK}
\begin{abstract}
\noindent
Using a new radio sample, 6C* designed to find radio galaxies at $z >
 4$ along with the complete 3CRR and 6CE sample we extend the radio
 galaxy $K-z$ relation to $z \sim 4.5$.  The 6C* $K-z$ data
 significantly improve delineation of the $K-z$ relation for radio
 galaxies at high redshift ($z >2$). Accounting for non-stellar
 contamination, and for correlations between radio luminosity and
 estimates of stellar mass, we find little support for previous claims that the
 underlying scatter in the stellar luminosity of radio galaxies
 increases significantly at $z >2$. This indicates that we are not probing
 into the formation epoch until at least $z\, \gtsim\, 3$.

\end{abstract}

\section*{Why radio galaxies?}
Radio galaxies provide the most direct method of investigating the
host galaxies of quasars if orientation based unified schemes are
correct. The nuclear light which dominates the optical/near-infrared
emission in quasars is obscured by the dusty torus in radio galaxies,
therefore difficulties surrounding the psf modelling and subtraction
are not required to determine the properties of the underlying host
galaxy. Unfortunately compiling samples of radio loud AGN is a long
process, because of the radio selection there is no intrinsic
optical magnitude limitation, making follow-up observations extremely
time consuming, especially when dealing with the faintest of these
objects.  However, low-frequency selected radio samples do now exist
with the completion of 3CRR (Laing, Riley \& Longair 1983) along with
6CE (Eales et al. 1997; Rawlings et al. 2001) and the filtered 6C*
sample (Blundell et al. 1998; Jarvis et al. 2001a; 2001b). We can now
use these radio samples to investigate the underlying stellar
populations through the radio galaxy $K-z$ Hubble diagram.

\section*{Previous radio samples and the $K-z$ Hubble diagram}
There has been much interest in the $K-z$ relation for radio galaxies
in the past decade. Dunlop \& Peacock (1993) using radio galaxies from
the 3CRR sample along with fainter radio sources from the Parkes
selected regions demonstrated that there exists a correlation between
radio luminosity and the $K-$band emission. Whether this is due to a
radio luminosity dependent contribution from a non-stellar source or
because the galaxies hosting the most powerful radio sources are
indeed more massive galaxies has yet to be resolved. Eales et
al. (1997) confirmed this result and also found that the dispersion in
the $K-$band magnitude from the fitted straight line increases with
redshift. This result, along with the departure to brighter magnitudes
of the sources at high redshift led Eales et al. to conclude that we
are beginning to probe the epoch of formation of these massive
galaxies. Using the highest redshift radio galaxies from ultra-steep
samples of radio sources van Breugel et
al. (1998) found that the near infrared colours of radio galaxies at
$z > 3$ are very blue, consistent with young stellar
populations. They also suggest that the size of the radio structure is
comparable with the size of the near infrared region, and the
alignment of this region with the radio structure is also more
pronounced at $z > 3$. Lacy et al. (2000) using the 7C-III sample
found evidence that the hosts of radio galaxies become more luminous with redshift
and are consistent with a passively evolving population which formed
at high redshift ($z > 3$). Thus, all of this work points to a radio
galaxy population which formed at high redshift and has undergone
simple passive evolution since. However, all of these studies were
made with only a few high-redshift ($z > 2$) sources. With the 6C*
sample we are now able to probe this high redshift regime with
increased numbers from samples with well-defined selection criteria.

\section*{The 6C* filtered sample}
\begin{figure*}[!ht]
{\hbox to \textwidth{\epsfxsize=0.95\textwidth
\epsfbox{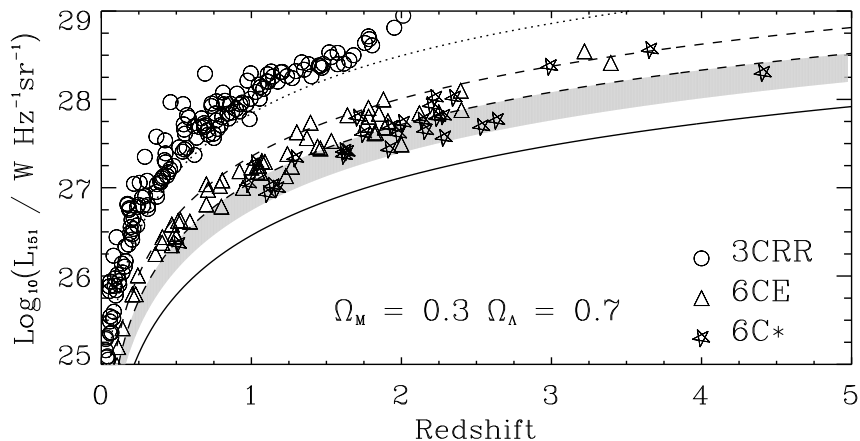}}} {\caption{\label{fig:pzplane} Rest-frame
151\,MHz luminosity ($L_{151}$) versus redshift $z$ plane for the 3CRR
(circles), 6CE (triangles) and 6C* (stars) samples.  The rest-frame
151\,MHz luminosity $L_{151}$ has been calculated according to a
polynomial fit to the radio spectrum (relevant radio data from
Blundell et al. 1998).  The curved lines show the lower flux-density
limit for the 3CRR sample (dotted line; Laing et al. 1983) and the
7CRS (solid line; Blundell et al. in prep; Willott et al. in
prep). The dashed lines correspond to the limits for the 6CE sample
(Rawlings et al. 2001) and the shaded region shows the 6C*
flux-density limits (all assuming a low-frequency radio spectral index
of 0.5). Note that the area between the 3CRR sources and 6CE sources
contains no sources, this is the area which corresponds to the absence
of a flux-density limited sample between the 6CE ($S_{151} \leq
3.93\:$Jy) and 3CRR ($S_{178} \geq 10.9\:$Jy) samples. The reason why
some of the sources lie very close to or below the flux-density limit
of the samples represented by the curved lines is because the spectral
indices lie very close to or below the assumed spectral index of the
curves of $\alpha = 0.5$.}}
\end{figure*}

The 6C* sample is a low-frequency radio sample ($0.96 {\rm Jy} \leq S_{151} \leq 2.00$\,Jy) which was originally designed to find radio sources at $z>4$ using filtering criteria based on the radio properties of steep spectral index and small angular size. The discovery of 6C*0140+326 at $z = 4.41$ (Rawlings et al. 1996) and 6C*0032+412 at $z = 3.66$ (Jarvis et al. 2001a) from a sample of just 30 objects showed that this filtering was indeed effective in finding high-redshift objects. Indeed, the median redshift of the 6C* sample is $z \sim 1.9$ whereas for complete samples at similar flux-density levels the median redshift is $z \sim 1.1$ (Willott et al. in prep.). We can now use this sample to push the radio galaxy $K-z$ diagram to high redshift ($z > 2$) where it has not yet been probed with any significant number of sources (e.g. Eales et al. 1997; van Breugel et al. 1998; Lacy et al. 2000). Fig.~\ref{fig:pzplane} shows the radio luminosity-redshift plane for the 3 samples used in this analysis.

\section*{Emission-line contamination}
The most-luminous sources at high redshift may be contaminated by the
bright optical emission lines redshifted into the infrared. This
is particularly true for sources in radio
flux-density limited samples. The high redshift sources in these
samples are inevitably some of the most luminous, and we also know
there is a strong correlation between low-frequency radio
luminosity and emission-line strength (e.g. Rawlings \& Saunders 1991; Willott et al. 1999; Jarvis et al. 2001a) which will increase the contribution to the measured $K-$band magnitudes from the emission-lines in the most radio luminous
sources.

\begin{figure}[!ht]
{\hbox to \textwidth{\epsfxsize=\textwidth \epsfbox{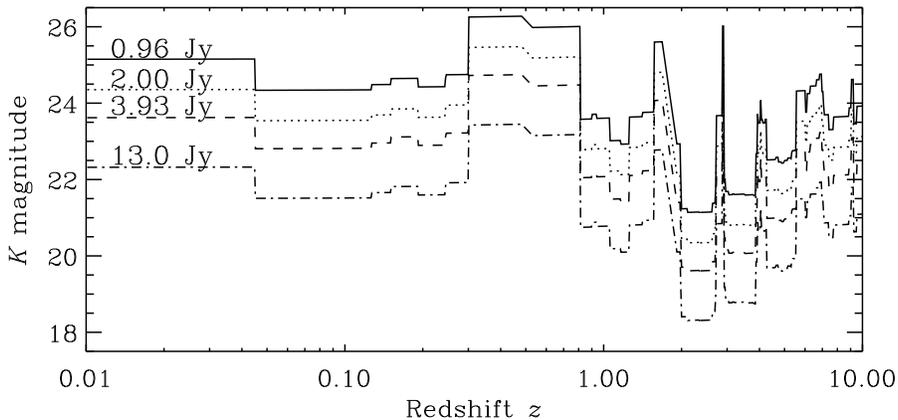} }}
{\caption{\label{fig:line_cont} Emission line contribution to the
$K-$band magnitudes for various radio flux-densities assuming the
power-law relation of $L_{\rm [OII]} \propto L_{151}^{1.00}$.}}
\end{figure}

To subtract this contribution we use the correlation between [OII]
emission-line luminosity $L_{\rm [OII]}$ and the low-frequency radio
luminosity $L_{151}$ from Willott (2000), where $L_{\rm [OII]} \propto
L_{151}^{1.00 \pm 0.04}$. Then by using the emission-line flux ratios
for radio galaxies (e.g. McCarthy 1993) we are able to determine the
contribution to the $K-$band magnitude from all of the other
emission-lines. This is illustrated in Fig.~\ref{fig:line_cont} where
the emission-line contamination to the $K-$band flux is shown for
various radio flux-density limits and a range of redshifts.

\section*{The $K-z$ relation}
\begin{figure}[!ht]
{\hbox to \textwidth{\epsfxsize=\textwidth \epsfbox{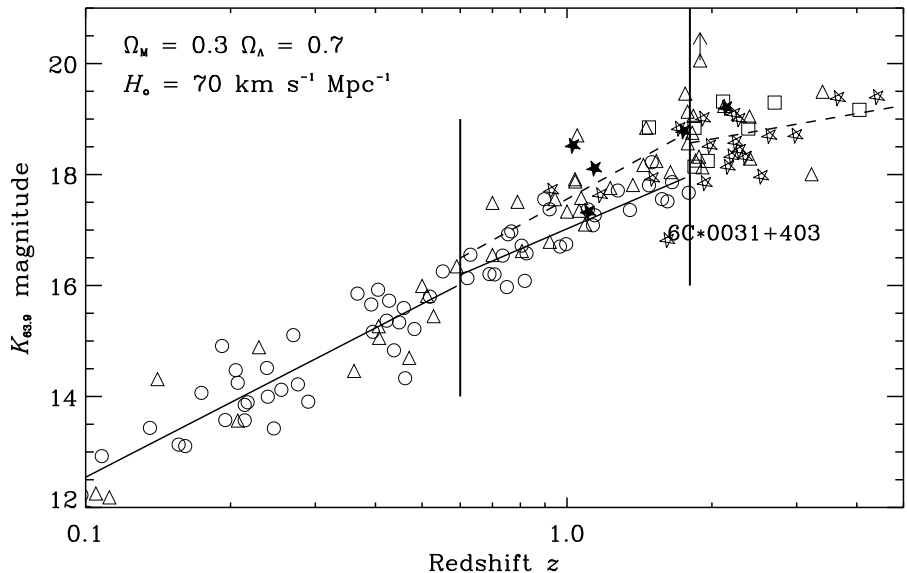} }}
{\caption{\label{fig:kzs} The $K-z$ Hubble diagram for radio galaxies
from the 3CRR (circles), 6CE (triangles), 6C* (stars) and 7C-III
(squares) samples. $K_{63.9}$ denotes the $K-$band magnitude within a comoving metric aperture of 63.9 kpc (c.f. Eales et al. 1997; Jarvis et al. 2001b).
The two vertical lines show the redshift above
which the alignment effect begins to be seen ($z = 0.6$) and the
higher redshift at which we chose to split the data beyond which
there are no 3CRR sources ($z = 1.8$). The solid lines are the fits
to the 3CRR data points at $z < 0.6$ and $0.6 < z < 1.8$. The dashed
line is the fit to the 6CE and 6C* sources at $0.6 < z < 1.8$ and $z >
1.8$. 6C*0031+403 probably has an AGN component contributing to the
$K-$band magnitude and is labelled. The filled
stars represent the objects in 6C* which do not yet have completely
secure redshifts.  }}
\end{figure}

Following the prescription of Eales et al. (1997) we split the data
into three redshift bins: $z < 0.6$, the redshift above which the
alignment effect is readily observed (e.g. McCarthy 1993); $0.6 \leq z
\leq 1.8$, the medium-redshift bin to compare 3CRR and 6CE/6C* sources
at the same redshift; and $z > 1.8$, the redshift above which there
are no 3CRR sources. A least-squares fit line to the $K-$band
magnitudes is plotted on the $K-z$ diagram for the three samples used
in our analysis (Fig~\ref{fig:kzs}). 
The main results from this plot
are:

(i) We confirm the previous results of Dunlop \& Peacock (1993)
and Eales et al. (1997) and find a correlation between radio
luminosity and near infrared magnitude. Sources from the new 6C* sample
occupy a similar range in $K-$band magnitudes as the 6CE sources,
which are a factor of two brighter in radio luminosity than the 6C*
sources. The 3CRR sources, which are a factor of six brighter than
the 6CE sources occupy $K-$band magnitudes approximately 0.7
magnitudes brighter than the 6CE/6C* sources at $0.6 <z < 1.8$ (A Mann-Whitney U-test shows that the distribution in magnitudes between the 6CE/6C* and 3CRR sources are different at $>99.9$\% level). 

(ii) We find that the dispersion in the $K-$band magnitudes for the
3CR sources decrease from the $z < 0.6$ bin ($\sigma = 0.52$) to the
$0.6 \leq z \leq 1.8$ bin ($\sigma = 0.36$). This may be because of a
non-stellar component linked to the radio luminosity becoming more
important at the higher redshifts as the sources are necessarily more
radio luminous than their low-redshift counterparts and such a luminosity dependent effect would reduce the observed dispersion. However, an
alternative explanation in which these extreme objects can only exist
in specific physical conditions also needs to be explored. It is not
inconceivable that the most luminous radio galaxies need some of the
narrowest set of conditions to form and exist, whereas the lower
luminosity radio galaxies may be able to form and exist in a broader
range of physical environments.

(iii) The dispersion in the $K-z$ diagram does not increase from $z
< 0.6$ ($\sigma = 0.57$) out to $z \sim 3$ ($\sigma = 0.51$) for
the sources in the 6CE and 6C* samples. This is in opposition to
results in which the dispersion was found to increase, which led
previous authors to conclude that we are probing the epoch of
formation at $z \sim 2$. The lack of an apparent increase in
dispersion toward high redshift may be due to the possibility outlined
above in which non-stellar emission may be contributing to the
$K-$band flux at these high redshifts, and thus
high luminosities. Alternatively, the lack of scatter may be informing
on the lack of ongoing star formation at these redshifts, and would
mean that the period of star formation has come to an end. If this
scenario is correct then it does fit in with other results concerning
the epoch at which these massive galaxies first formed. First,
Archibald et al. (2001) have found that the dust mass in radio
galaxies appears to increase with redshift, at least out to $z \sim 3$, thus
implying that the majority of star-formation activity in these
galaxies is occurring at high redshift. Second, the discovery of six
extremely red objects at $1 < z < 2$ in the 7C redshift survey Willott
et al. (2001; these proceedings) with inferred ages of a few Gyrs
implies that the bulk of their stellar population formed at $z \simeq
5$.



\begin{chapthebibliography}{<widest bib entry>}

\bibitem[Archibald et al. 2001]{arch01}
Archibald E.N., et al., 2001, MNRAS, in press (astro-ph/0002083)

\bibitem[Blundell et al. 1998]{blun98}
Blundell K.M., et al., 1998, MNRAS, 295, 265

\bibitem[Dunlop \& Peacock 1993]{DP93}
Dunlop J.S., Peacock J.A., 1993, MNRAS, 263, 936

\bibitem[Eales et al. 1997]{Eea97}
Eales S.A., Rawlings S., Law-Green D., Cotter G., Lacy M., 1997, MNRAS, 291, 593

\bibitem[Jarvis et al. 2001a]{Jea01a}
Jarvis M.J., et al., 2001a, MNRAS, submitted

\bibitem[Jarvis et al. 2001b]{Jea01b}
Jarvis M.J., et al., 2001b, MNRAS, submitted

\bibitem[Lacy, Bunker \& Ridgway 2000]{lbr00}
Lacy M., Bunker A.J., Ridgway S.E., 2000, AJ, 120, 68

\bibitem[Laing, Riley \& Longair 1983]{lrl}
Laing R.A., Riley J.M., Longair M.S., 1983, MNRAS, 204, 151

\bibitem[McCarthy 1993]{mcc93}
McCarthy P.J., 1993, ARAA, 31, 639

\bibitem[Rawlings, Eales \& Lacy 2001]{rel01}
Rawlings S., Eales S.A., Lacy M., 2001, MNRAS, 322, 523

\bibitem[Rawlings et al. 1996]{raw96}
Rawlings S., et al., 1996, Nature, 383, 502

\bibitem[Rawlings \& Saunders 1991]{rs91}
Rawlings S. \& Saunders R., 1991, Nature, 349, 138

\bibitem[van Breugel et al. 1998]{vB98}
van Breugel W.J.M., et al., 1998, ApJ, 502, 614

\bibitem[Willott 2000]{cjwconf}
Willott C.J., 2000, to appear in Proc. "AGN in their Cosmic Environment", Eds. B. Rocca-Volmerange \& H. Sol, EDPS Conf. Series (astro-ph/0007467)

\bibitem[Willott, Rawlings \& Blundell 2001]{cjwero}
Willott C.J., Rawlings S., Blundell K.M., 2001, MNRAS, in press (astro-ph/0011082)

\bibitem[Willott et al. 1999]{cjwemline}
Willott C.J., Rawlings S., Blundell K.M., Lacy M., 1999, MNRAS, 309, 1017

\end{chapthebibliography}
\end{document}